
\input phyklo
\input tables
\input defj
\input epsf

\hsize=16.2cm
\vsize=22.5cm
\hoffset=0pt \voffset=0.5cm
\overfullrule=0pt
\def\PRD#1 {{\it Phys. Rev.} {\bf D#1}\ }

\def\chapter#1{\par \penalty-300 \vskip\chapterskip
   \spacecheck\chapterminspace
   \chapterreset \leftline{\bf\chapterlabel \ #1}}
\def\section#1{\par \ifnum\the\lastpenalty=30000\else
   \penalty-200\vskip\sectionskip \spacecheck\sectionminspace\fi
   \wlog{\string\section\ \chapterlabel \the\sectionnumber}
   \global\advance\sectionnumber by 1  \noindent
   {\bf\chapterlabel \sectionlabel #1}\par  }

\def\DAF{DA$\Phi$NE}

\def\eps{\epsilon    }
\def\epoe{$\epsilon'/\epsilon$}

\def\rep{\ifm{{\rm Re}(\eps'/\eps)}}
\def\imp{\ifm{{\rm Im}(\eps'/\eps)}}
\FIG\figa
\FIG\figc
\FIG\figd
\FIG\fige
\FIG\figf
\FIG\figg
\FIG\figh
\FIG\figha
\FIG\figi
\FIG\figj


\parindent=1cm
\baselineskip=16pt
\vbox to 5pt{\vfill}\par
\vbox to 1.6 cm{\fourteenpoint
\vfill
\centerline{\bf PHYSICS AT \DAF}}

\twelvepoint
\vglue 7pt
\centerline{\bf  Paula J. Franzini}
\centerline{Paul Scherrer Institute, CH-5232 Villigen PSI, Switzerland}

\chapter{Introduction}
In this talk, I will give a brief description of the $\phi$ factory
\DAF\ at Frascati, explaining why a  $\phi$ factory is an interesting
place to do new physics, and then discuss the physics that can be done at
\DAF.  Since Ulf Meissner has already told you about $K$ decays at
\DAF\ and their relevance to CHPT (Chiral Perturbation Theory), and
Pierluigi Campana will cover some of the other \DAF\ physics topics
such as $\gamma\gamma$ physics, meson spectroscopy and $\Delta I=
{1\over 2}$ violation in hypernuclei, I will concentrate on CP violation
as it can be studied at \DAF.  This is, after all, the {\it raison
d'\^etre} of \DAF.  I start with a brief general introduction to CP
violation in the $K\bar K$ system, and the distinction between
mass-mixing CP violation ($\epsilon$) and {\it intrinsic} CP violation
(\epoe).  After presenting a summary of \epoe\ measurements up to
now, and briefly discussing the theory of \epoe\ (the so-called
`penguins'), I will cover the particularities of measuring
\epoe\ at a $\phi$ factory, such as ${\it tagging}$ and
{\it interferometry}.  Finally, I will say a few words about searching
for CP violation in modes where it has never before been seen.
I will end my talk with a list of other physics topics at \DAF, and
rare decay branching ratio limits that can be achieved there, just
to give a flavor of what else can be done.
\REF\jlf{J. Lee-Franzini,
{\it Les Rencontres de Physique de la Valle\'e d'Aoste,
La Thuile, Italy, March 8--14  1992}, M. Greco Ed., p. 349.}
\REF\meee{P. J. Franzini,
{\it Les Rencontres de Physique de la Valle\'e d'Aoste,
La Thuile, Italy, March 3--9  1991}, M. Greco Ed., p. 257.}
For a starting point for more information about physics at \DAF, see
Ref. \jlf; for \epoe\ theory and  history see Ref. \meee.

\chapter{Why a $\phi$ factory? What is \DAF?}
First of all, why do we want a $\phi$ factory?  Because, essentially,
a $\phi$ factory is a $K$ factory.  And these kaons are not just any kaons,
but kaons in a well-defined quantum-mechanical and kinematic state.
This,
as we shall see, is very important.  The $\phi(1019)$ is the lowest lying
$J^{PC}=1^{--}$ bound state of a strange quark and a strange anti-quark.
\DAF\ is an $e^+e^-$ collider optimized to run at the center-of-mass (c.m.)
energy $M_\phi$, due to deliver a luminosity ${\cal L}= 10^{32}
{\rm cm}^{-2} {\rm s}^{-1}$ in 1996.  The cross section for $e^+ e^-
\rightarrow \phi$ at the $\phi$ resonance peak is about 5 $\mu$b, meaning
that at the eventual
target luminosity ${\cal L}= 10^{33}{\rm cm}^{-2} {\rm s}^{-1}$,
5000 $\phi$'s are produced per second.  Using the canonical high energy
physics definition of one `machine year'=$10^7$ seconds, giving a leeway
of about a factor of $\pi$ to account for the various integrated luminosity
degradation factors (e.g. machine study, maintenance and down periods;
detector down periods; and peak versus average luminosity), this means
$5\times 10^{10}$ $\phi$'s per year!

The $\phi(1019)$, with a mass $M_\phi=1019.412\pm 0.008$ MeV, total
width $\Gamma=4.41$ MeV, and leptonic width $\Gamma_{ee}=1.37$ keV,
decays into the modes given in Table 1.
 Here BR is the
branching ratio, in percent; $\beta_K$ is the $\beta$ (= velocity$/c$)
of the kaon; $\gamma\beta c \tau ]_K$ its mean path length in centimeters;
and $p_{max}$ is the momentum of the resultant particles in MeV (maximum
momentum if there are three particles).  The last column gives the
resultant number of such decays in the canonical year, demonstrating
that \DAF\ is indeed a factory of neutral kaons in a well prepared
quantum state, and
 of charged $K$ pairs, as well as of $\rho$'s, $\eta$'s
(and rarer $\eta'$'s).  These numbers of kaons will have to be reduced
by the {\it tagging efficiencies} of about 30--80$\%$, depending on
kaon species, in order to get the number of useful, well-identified
kaons (see Sec. 6).
The high luminosity of \DAF\ will also allow
measurements of rare $\phi$ radiative decays (see Sec. 9).

\vskip .2in
\begintable
 Mode     | BR  | $\beta_K$ | $\gam\beta c\tau]_K$ | $p_{\rm max}$
| \#
\nrneg{4pt}
            |  \%  |          | cm     | MeV/c  | \cr
 $K^+K^-$   | 49   |  0.249   |  95.4  | 127 | 2.5$\times 10^{10}$ pairs\cr
   \ksl\    | 34   |  0.216   | 343.8  | 110 | 1.5$\times 10^{10}$ \cr
 $\rho\pi$  | 13   |    --  |  --   | 182 |6$\times 10^{9}$ \cr
 \pic\po\   | 2    |  --    |  --   | 462 |1$\times 10^{9}$ \cr
 $\eta\gam$ | 1.3  |  --    |  --   | 362 |6$\times 10^{8}$\cr
 other     | \ab1  |  --    |  --   | -- |5$\times 10^{8}$ \endtable
\centerline{
{\bf Table 1.} $\phi$ decays. }
\vskip .2in

The main goal of \DAF\ is to measure direct CP violation (\epoe)
to accuracies of about $1\times 10^{-4}$ by observing $K_{L,S}
\rightarrow \pi^0 \pi^0$, $\pi^+\pi^-$.  In general, also, \DAF\
is exciting not only because it is the first $\phi$ factory, but
because it is the next {\it new} particle physics accelerator
(which means new results!).  This statement may be somewhat
contestable, depending on what one calls particle physics and
what one calls {\it new}, so a more unambiguous claim is that it
will be the {\it first} machine of the factory era, the first of a
new generation of super-high luminosity (in the
$10^{33}{\rm cm}^{-2} {\rm s}^{-1}$ range, to be contrasted with
$10^{31}{\rm cm}^{-2} {\rm s}^{-1}$ for existing machines)
$e^+ e^-$ colliders, designed to stay within a narrow range of
energy and produce a large number of a given particle (or family
of particles).

What makes a factory?  The luminosity of a collider is given by the
following formula,
$${\cal L} = f n {N_1 N_2 \over A},
\eqn\zi$$
where $f$ is the revolution frequency, $n$ is the number of bunches,
$N_i$ is the number of particles per bunch for each species of particle,
and $A$ is the area of the beams (for this formula to be valid, the
beams must be 100$\%$ overlapping).  Thus, many bunches of many particles,
going around at a high frequency, focussed tightly into beams of a
very small cross section, produce a high luminosity.  Nonetheless,
whichever of these parameters are modified to produce a larger luminosity,
without radically new technology, the luminosity in a single
ring machine will be limited by
{\it beam-beam interactions} to be about that of current machines
($10^{31}{\rm cm}^{-2} {\rm s}^{-1}$).  Synchrotron oscillations
`shake' up the beams and destroy the small bunch size needed for high
luminosity; bunches containing more particles
lead to stronger oscillations.  Multiple
bunches in a single ring do not help; each bunch `sees' $n$ of its
counterparts and gets successively more and more perturbed.

 The
solution generally found in `factories' is to have two separate rings
(hence the DA in \DAF, for Double Annular),
which cross each other at a small but non-zero {\it crossing angle}
(20 to 30 mrad for \DAF).  This crossing angle is needed, even though
head-on collisions are less disruptive to the beam, because if the
two beams were parallel even for a few meters, each bunch would then
pass several of its counterparts in a small machine like \DAF, where there
will be more than one bunch per meter.  The \DAF\ main rings are 98 meters
in perimeter, in a roughly rectangular shape, of 32 by 23 meters.
The machine will commence operation with 30 bunches of $9\times 10^{10}$
$e^\pm$ per bunch, yielding a luminosity of about $1\times
10^{32}{\rm cm}^{-2} {\rm s}^{-1}$.  At this level already, that means
about the same number of particles in a 98 m ring as LEP currently has
in its 27 km ring.  \DAF\ will then go to 120 bunches; this is the maximum
possible
number of bunches at the planned RF frequency of 368.25 MHz.  Fine-tuning
of the machine parameters is then expected to bring the luminosity
to its final level of $1\times 10^{33}{\rm cm}^{-2} {\rm s}^{-1}$.
The bunches are long and flat: 3 cm long in the beam direction,
2-3 mm in the horizontal direction perpendicular to the beam, and only
20 $\mu$m thick in the vertical direction.   The beams cross at
a small angle
in the horizontal plane.  Historically, at DORIS, synchro-betatron
oscillations were a problem with vertical crossings;
with horizontal crossings, the particles of one flat bunch will be well
embedded in those of the other flat bunch, and thus not disturb each
other as they do when they are in different planes.

The machine complex of \DAF\ is shown in Fig.~{\figa}.
There will be
two interaction areas, for the two experiments KLOE (particle physics)
and FINUDA (nuclear physics), which will be discussed by Pierluigi
Campana.  There is as well the possibility for medical research and
so on with the ultraviolet and x-ray  beams of the {\DAF}-L(ight)
facility.   The \DAF\ beam energy will be
$0.51$ GeV $\pm$ $0.4$ MeV, the same as the injection energy, which
means there will be no acceleration in the ring in the normal mode of
running, on the $\phi$ resonance.  The $e^\pm$ are accelerated in the
{\it linac}, a linear accelerator consisting of two pieces, one that imparts
a maximum of 250 MeV, the other 550 MeV.  The electrons thus could in
principle be accelerated to a maximum of 800 MeV.  The positrons are
created in the first segment and accelerated in the second.  The
beams are then stored and `cooled' (meaning that the momentum spread
-- and as a result, the position spread -- is decreased) in the
accumulator, a small ring, before injection into the main rings.
\endpage

\epsffile[30 -50  550 550]{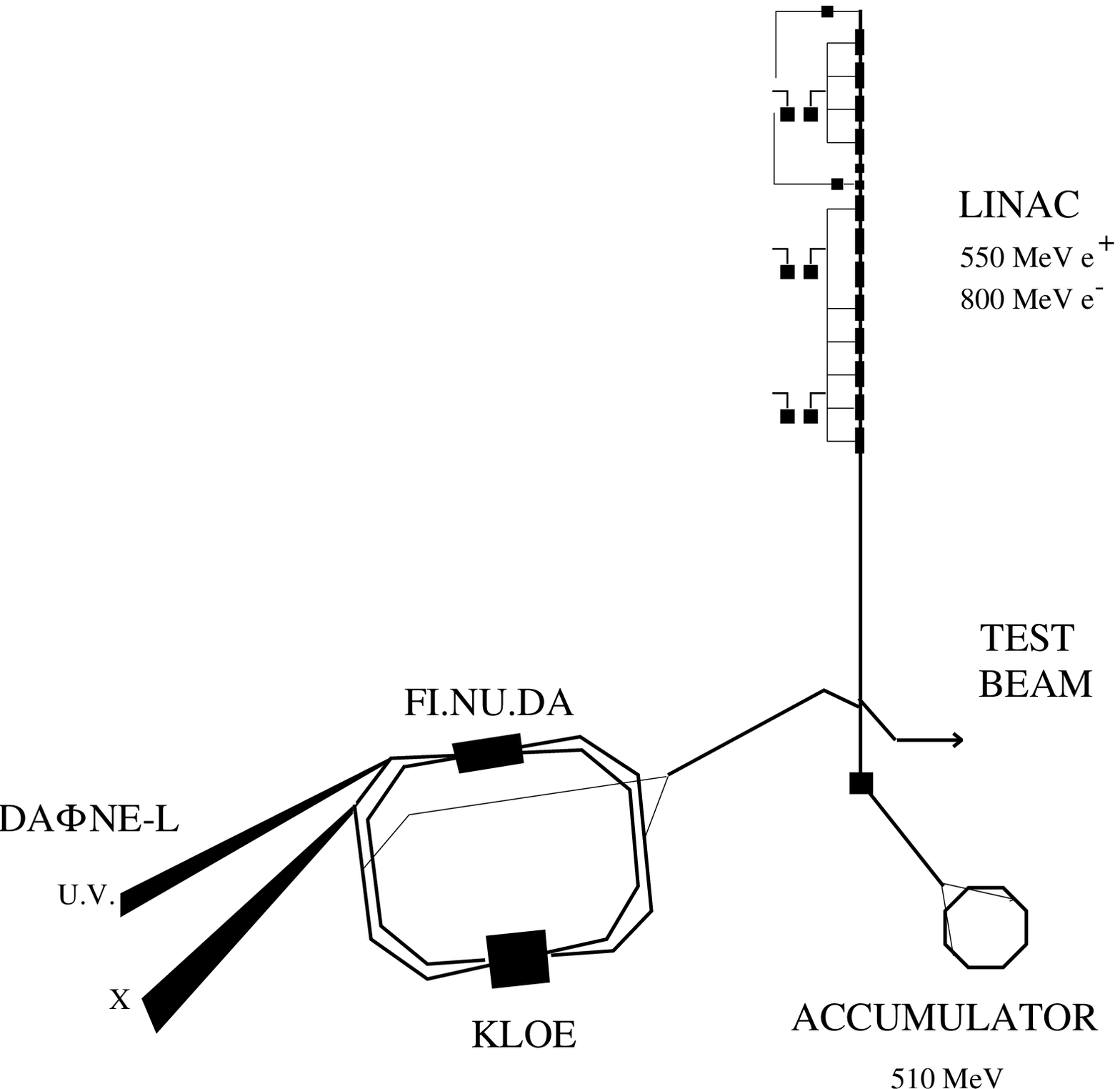}

\centerline{{\bf Figure~{\figa}.} The \DAF\ machine complex.}

\endpage
In Fig.~{\figc}, the luminosity versus center of mass energy
($E_{\rm c.m.}$) for
existing single ring $e^+ e^-$ colliders (the maximums that have been
achieved) are contrasted with the expected range for factories-to-be.
Beauty factories have been approved for construction at SLAC
and at KEK, while
the $\tau$-charm factory is  so far only intermittently
under consideration.

\epsffile[0 0  390 390]{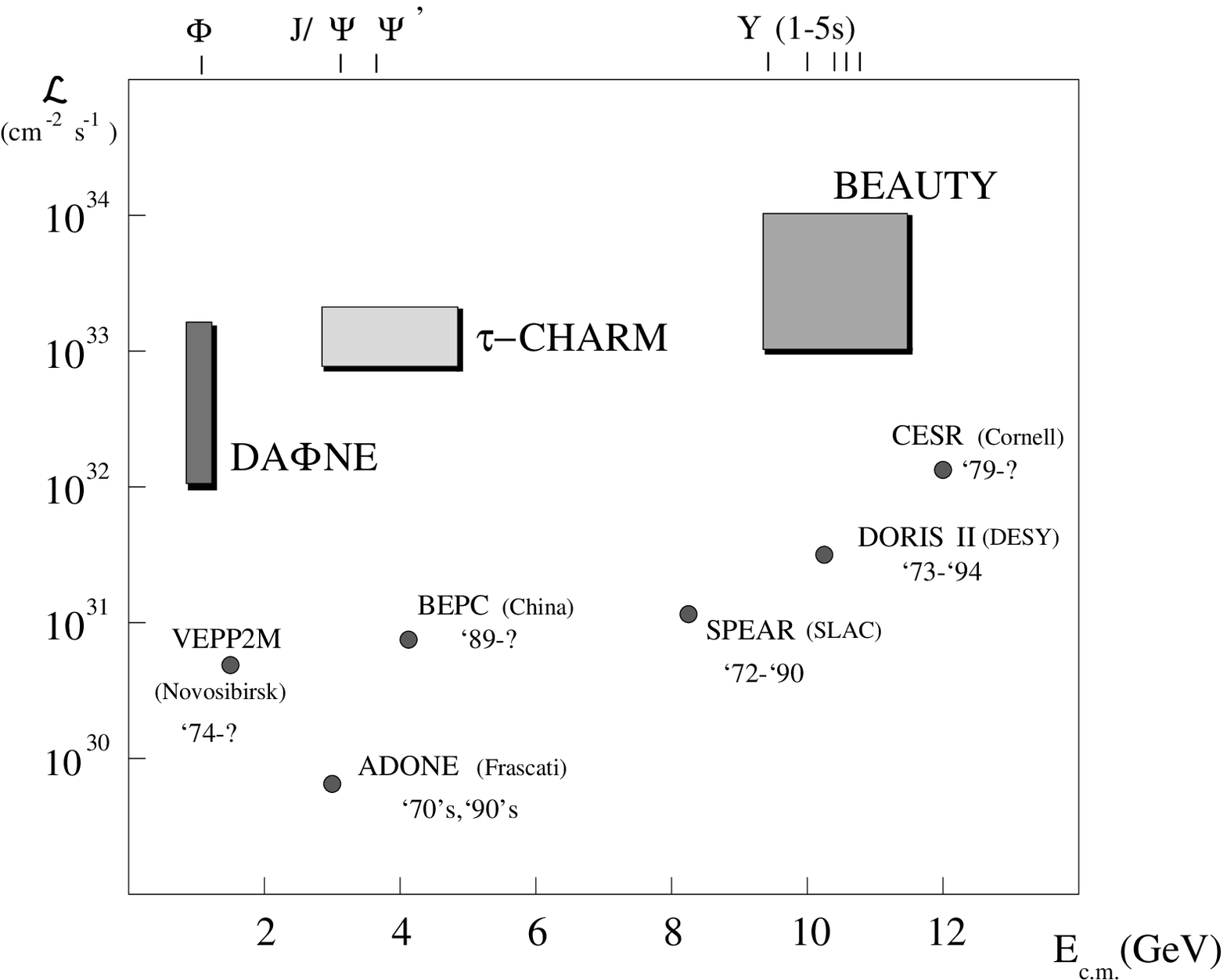}

{\bf Figure~{\figc}.} Luminosity versus center of mass energy for
existing single ring $e^+e^-$ colliders (dots) contrasted with
the expected ranges for factories-to-be.  At the top of the figure
are indicated the energies of the various resonances to be studied
at these factories.
\vskip .2in

\chapter{CP violation in the $K\bar K$ system}
Kaons contain a strange quark ($s$) or antiquark ($\bar s$), and are
composed of two isospin doublets: $K^0=d \bar s$, $K^+= u\bar s$, the
$S=+1$ doublet, and $\bar K^0=\bar d  s$, $K^-= \bar u s$, $S=-1$.
Kaons are {\it produced} by strangeness conserving {\it strong}
interactions.  {\it Weak} interactions do not conserve $S$, and
mix $K^0$ and $\bar K^0$ via the transition $K^0\leftrightarrow
2\pi \leftrightarrow \bar K^0$, represented at the quark level by the {\it
box} diagrams in Fig.~{\figd}.  C violation allows these transitions
to occur; CP need not be violated.
Under CP, the $K^0$ and $\bar K^0$ transform into each other:
$$CP \ket{K^0} =  \ket{\bar{K^0}} \ \ \ \ \
CP \ket{\bar {K^0}} =  \ket{K^0}.
\eqn\zii$$
We can then write down the linear combinations of these states that
are CP eigenstates; if CP is conserved, these will be the physical
eigenstates as well:
$$ \ket{K_1} = {{ \ket{ K^0}} + \ket{\bar {K^0}} \over \sqrt{2} } \ \ \ \ \
 \ket{K_2} ={ { \ket {K^0}} - \ket{\bar {K^0}} \over \sqrt{2} }.
\eqn\ziii$$
$\ket{K_1}$ is thus CP even (CP$\ket{K_1}$ = $+\ket{K_1}$ ) and
must decay to two pions, a CP even state.  $\ket{K_2}$ is CP
odd and must decay to three pions.

\vskip .5cm
\hskip .5cm $d$        \hskip 2.1cm $u$,$c$, $t$ \hskip 2.1cm $\bar d$
\hskip .6cm
\hskip .5cm $d$        \hskip 2.2cm $W$ \hskip 2.3cm $\bar d$
\vskip .55cm
\hskip 3.2cm $W$ \hskip 6.2cm $u$,$c$,$t$
\vskip .15cm
\epsffile[-50 3  300 8]{box.ps}

\hskip .5cm $\bar s$        \hskip 2.1cm $\bar u$,$\bar c$, $\bar t$
\hskip 2.1cm $s$
\hskip .6cm
\hskip .5cm $\bar s$        \hskip 2.2cm $W$ \hskip 2.4cm $s$

\centerline
{{\bf Figure~{\figd}.} The box diagrams mediating the $K^0-\bar K^0$
transition.}
\vskip .2in

Thus, if one has a beam of kaons, one
will see several two pion decays, close together, near the origin
of the beam; then several meters down the beamline, a spread-out
sprinkling of three pion decays.    However, in 1964, Cronin and
Fitch\Ref\CF{J. H. Christenson, J. W. Cronin, V. L. Fitch and
R. Turlay, Phys. Rev. Lett. {\bf 13} (1964) 138.}
observed {\it long lived} kaons decaying to two pions!
This was the first evidence for CP violation.  Physical states are
thus no longer CP eigenstates.  We instead have physical states
called $K_S$, the short-lived state, and $K_L$, the long-lived state.
$K_S$ will have a small admixture of CP-odd $K_2$, while $K_L$ has
a small admixture of CP-even $K_1$, in other words:
$$ \ket{K_S} = { \ket{K_1} + \epsilon \ket{K_2} \over
\sqrt{1 + |\epsilon|^2} } = {1+\epsilon \over\sqrt{1 + |\epsilon|^2} }
\ket{K^0} +  {1-\epsilon \over\sqrt{1 + |\epsilon|^2} } \ket{\bar K^0}
\eqn\ziv$$
$$ \ket{K_L} = { \ket{K_2} + \epsilon \ket{K_1} \over
\sqrt{1 + |\epsilon|^2} } = {1+\epsilon \over\sqrt{1 + |\epsilon|^2} }
\ket{K^0} -  {1-\epsilon \over\sqrt{1 + |\epsilon|^2} } \ket{\bar K^0}.
\eqn\zv$$
In this picture, $K_L$ can decay to two pions simply because it has
a small component of $K_1$.  If $K_2$ has no {\it intrinsic} CP violation,
i.e.,
$$\bra{K_2} H_W \ket{\pi\pi} = 0,
\eqn\zvi$$
then
$$ { \bra{K_L} H_W \ket{\pi\pi} \over \bra{K_S} H_W \ket{\pi\pi}} =
 { \epsilon \bra{K_1} H_W \ket{\pi\pi} \over \bra{K_1} H_W \ket{\pi\pi}} =
\epsilon.
\eqn\zvii$$
CP violation
was discovered in 1964, implying a non-vanishing value for $\epsilon$, which
has been quite well-measured
for quite some time, with value
$$|\eps| = 2.259 \pm 0.018 \times 10 ^{-3}.
\eqn\zviii$$
The question now is, is there {\it intrinsic} CP violation?  Thirty years
after the discovery of CP violation, this question still remains
unanswered.

\chapter{Measuring direct CP violation in the processes
$K_{L,S}\rightarrow\pi\pi$ }
A state consisting of a pair of pions must have an isospin of zero or
two.  The Clebsch-Gordan table for the combination of two isopin one
states is the following.

\vskip .2in
\begintable
 $1\times$ & 1    | 2 | 1 | 0 \crnorule
           &       | 0 | 0 | 0 \cr
$\ 1$ & $-1$         | $1/\sqrt{6}$ | $1/\sqrt{2}$ | $1/\sqrt{3}$ \crnorule
 $\ 0$ & $\ 0$            | $\sqrt{2/3}$ | 0 | $-1/\sqrt{3}$ \crnorule
 $-1$ & $\ 1$         | $1/\sqrt{6}$ | $-1/\sqrt{2}$ | $1/\sqrt{3}$ \endtable

\noindent Thus we have, for appropriately symmetrized dipion states,
$$ {1\over \sqrt{2} } \left( \ket{\pi^+ \pi^-}+\ket{\pi^- \pi^+}
\right) = \sqrt{2\over 3} \ket{I=0} +  \sqrt{1\over 3} \ket{I=2}
\eqn\zix$$
$$  \ket{\pi^0 \pi^0} = -\sqrt{1\over 3} \ket{I=0}
 +  \sqrt{2\over 3} \ket{I=2}.
\eqn\zx$$
Let us define the amplitudes
$$\bra{\pi\pi I=0} H \ket{K^0} \equiv A_0 e^{i\delta_0}
\ \ \ \ \  \left(
  \bra{\pi\pi I=0} H \ket{\bar K^0} = - A^*_0 e^{i\delta_0}\right)
\eqn\zxi$$
$$\bra{\pi\pi I=2} H \ket{K^0} \equiv A_2 e^{i\delta_2}
\ \ \ \ \  \left(
  \bra{\pi\pi I=2} H \ket{\bar K^0} = - A^*_2 e^{i\delta_2}\right).
\eqn\zxii$$
Recall eq. \zii.  In the CP conservation limit, $A_{0,2}=
A^*_{0,2}$, i.e., the amplitudes for $K^0$ and $\bar K^0$ differ
only by a sign, and have no phase difference.  The phases
$\delta_{0,2}$ are the phases that $K^0$ and $\bar K^0$ have in
common, while an imaginary part to $A_{0,2}$ indicates a CP
violating phase {\it difference} between $K^0$ and $\bar K^0$.
Actually, there is one unphysical phase (since we can only measure
intensities, not amplitudes), so it is the difference in phase
between $A_{0}$ and $A_{2}$
that indicates direct CP violation.

One common phase convention is to take $A_0$ real.  One can then
derive (an exercise left for the reader) from eqs. \ziv, \zv,
and \zix$-$\zxii,
dropping higher order terms (and bearing in mind that
$\epsilon\ll 1$, $\epsilon'\ll\epsilon$ and $A_2\ll A_0$),
the relations
$$\eta^{\pm} \equiv {\bra{\pi^+\pi^-} H \ket{K_L} \over
                     \bra{\pi^+\pi^-} H \ket{K_S} } \approx
\epsilon + \epsilon'
\ \ \ \ \ \ \ \ \ \
\eta^{00} \equiv {\bra{\pi^0\pi^0} H \ket{K_L} \over
                     \bra{\pi^0\pi^0} H \ket{K_S} } \approx
\epsilon - 2\epsilon'
\eqn\zxiii$$
where $\epsilon'$ is given by the expression
$$\epsilon' = {i\left( {\rm Im}
A_2 \right) e^{i(\delta_2 - \delta_0)} \over \sqrt{2} A_0}.
\eqn\zxiv$$
Note that here $\bra{\pi^+\pi^-}$ is taken to imply a symmetrized
dipion state.

Thus, the following three statements are  equivalent:
\point $\epsilon'\neq 0$;
\point $A_2$ is complex in the basis in which $A_0$ is real;
\point there is {\it direct} CP violation, i.e., there is
CP violation in the decay $K_2\rightarrow \pi\pi$.

Moreover, from the expressions for  $\eta^{\pm}$ and $\eta^{00}$,
we see that ${\epsilon'/\epsilon}$ can be measured via
the so-called {\it double ratio}
$$ {N(K_L\rightarrow \pi^+ \pi^-) \over N(K_S\rightarrow \pi^+ \pi^-) }
\Bigg/
{N(K_L\rightarrow \pi^0 \pi^0) \over N(K_S\rightarrow \pi^0 \pi^0) }
\approx { |\epsilon +\epsilon'|^2 \over  |\epsilon -2 \epsilon'|^2 }
\approx 1 + 6 {\rm Re} {\epsilon'\over \epsilon}.
\eqn\zxv$$
Other similar ways of measuring ${\epsilon'/ \epsilon}$
are ratios such as the following:
$${N(\pi^+ \pi^- \pi^+ \pi^-) \over N(\pi^0 \pi^0 \pi^0 \pi^0) }
\times \left( BR(K_S\rightarrow \pi^0 \pi^0) \over
 BR(K_S\rightarrow  \pi^+ \pi^-) \right)^2
\approx 1 + 6 {\rm Re} {\epsilon'\over \epsilon}
\eqn\zxvi$$
and
$${N(\pi^+ \pi^- \pi^+ \pi^-) \over N(\pi^+ \pi^- \pi^0 \pi^0) }
\times \left( BR(K_S\rightarrow \pi^0 \pi^0) \over
 BR(K_S\rightarrow  \pi^+ \pi^-) \right)
\approx 1 + 3 {\rm Re} {\epsilon'\over \epsilon}.
\eqn\zxvii$$
The statistical uncertainties on these three and similar ratios
will all be the same, but the systematic errors may differ.  Thus
different ratios may be optimal for different experiments, or
different ratios may serve as valuable counterchecks in a given
experiment.

Currently, the most precise determinations of
$ {\epsilon'/\epsilon}$ are those of the experiments
NA31\Ref\nathr{G. D. Barr, Phys. Lett. {\bf 317B} (1993) 233.}
at CERN, and E731\Ref\esev{L. K. Gibbons {\it et al.}, Phys.
Rev. Lett. 70 (1993) 1203.}
at Fermilab.  They find
$$ \eqalign {
{\rm Re} (\epsilon'/\epsilon) =& 2.3 \pm 0.65 \times 10^{-3}
\ \ ({\rm NA31}) \cr
& 0.74 \pm 0.60 \times 10^{-3}\ \  ({\rm E731}),\cr}
\eqn\zxviii$$
the one consistent with  ${\epsilon'/ \epsilon}\neq 0$,
the other consistent with ${\epsilon'/ \epsilon} =0$, and
yet both consistent with each other.
These experiments create
high-energy $K$ beams (of the order of 100 GeV) by collisions
of high-energy protons (450 GeV for NA31, 800 GeV for E731)
with a fixed target.  NA31 alternates operation with a $K_L$
beam and with a $K_S$ beam, while E731 uses simultaneously
two $K_L$ beams, one of which, by virtue of
regenerators,\foot{An initially pure $K^0$ beam
quickly becomes essentially
a $K_L$ beam as the $K_S$ part decays.  However, a $K_L$ beam
passing through {\it matter} will lose more $\bar K^0$ than $K^0$,
as $\bar K^0$ reacts more in nuclear collisions
(since $\bar K^0$ contains $\bar d$, a light {\it anti}quark).
This passage through matter thus {\it regenerates}
the $K_S$ component.}
produces the required $K_S$'s.  In a few years, these two experiments
also hope to come out with measurements of ${\epsilon'/ \epsilon}$
at the same accuracies that \DAF\ is aiming at, namely
at the $10^{-4}$ level, with clearly very different, and thus
very complementary, systematic effects.

Fig.~{\fige}
 shows the evolution of the measured value of \epoe\ over the
last two decades.
\epsffile[0 0  320 320]{hist.eps}

{\bf Figure~{\fige}.} Evolution of experimental measurements
and theoretical estimates of Re \epoe\ (in units of $10^{-3}$)
over the last two decades.  See text for detailed description.
\vskip .2in
\noindent Two decades is a convenient cutoff, since many
of the results before 1970 tend to be based on measurements of
 $\eta^{\pm}$ and $\eta^{00}$ by separate experiments.
The results here are, in chronological order:
1972, Holder et al.,\Ref\Holder{M. Holder et al.,
Phys. Lett. {\bf 40B} (1972) 141.} CERN;
1972, Banner et al.,\Ref\Banner{M. Banner et al., Phys. Rev.
Lett. {\bf 28} (1972) 1597.} Brookhaven/Princeton;
1979, Christenson et al., \Ref\Chri{J. H. Christenson et al.,
Phys. Rev. Lett. {\bf 43} (1979) 1209.} Brookhaven/NYU;
1985, Black et al.,\Ref\Black{J. K. Black et al., Phys. Rev.
Lett. {\bf 54} (1985) 1628.} Brookhaven/Yale;
1985, E731;\Ref\fermia{R. H. Bernstein et al., Phys. Rev. Lett.
{\bf 54} (1985) 1631.}
1988, E731;\Ref\fermib{M. Woods et al., Phys. Rev. Lett.
{\bf 60} (1988) 1695.}
1988, NA31;\Ref\cerna{H. Burkhardt  et al., Phys. Lett.
{\bf 206B} (1988) 169.}
1990, E731\Ref\fermic{J. R. Patterson  et al., Phys. Rev. Lett.
{\bf 64} (1990) 1491.}
and finally the two most recent measurements, quoted above, whose
errors are so small that they are denoted by the size of the points
alone.\foot{That there are less points shown for NA31 than E731
does not mean that NA31 did not publish any results between
1988 and 1993, but as their results did not change dramatically
I have not shown their intermediate results.}
The errors of the next generation of experiments, nearly
ten times smaller than these, will require a new graph with a
different scale!  In this figure is also shown the evolution of
theoretical predictions for \epoe, which will be described in
the next section.  These predictions are all within
the framework of the Standard Model.

\chapter{Theory of \epoe\ in a nutshell}
We have seen in the previous sections that if CP violation is
due only to the {\it mixing} of $K_1$ and $K_2$, $\epsilon'=0$.
If there is CP violation in the {\it decay} of $K_2$ as well,
$\epsilon'\neq 0$.  Mixing is described by the box diagrams
in Fig.~{\figd}.  The decay is described by analogous box
diagrams, and
the so-called penguin
diagram, shown in Fig.~{\figf} (which I have taken pains to draw
as much like a penguin as possible --- nonetheless...).

\vskip .2in
\hskip2.3in\vbox{\hsize=3.52in
In the Standard Model (SM) {\it both} penguin diagrams and
box diagrams are proportional to $\sin \delta$, where
$\delta$ is the one non-trivial (and unknown)
Cabibbo-Kobayashi-Maskawa (CKM) phase.  In
1973, Kobayashi and
Maskawa\Ref\KM{M. Kobayashi and K. Maskawa, Prog. Theor. Phys.
{\bf 49} (1973)652.} showed explicitly that a three generation model
can account for CP violation.
They proved that with a two generation model, one has no
non-trivial phase, and thus no CP violation (in the SM).
Thus, in the SM, barring accidental cancellations
that we will see later,
both $\epsilon$ and \epoe\ should be non-zero.}

\epsffile[-20 -20  -15 -15]{penguin.eps}

\noindent{\bf Figure~{\figf}.} The gluonic penguin
\vskip .2in

\noindent In contrast, in the `superweak' model of
Wolfenstein\Ref\Wolfcp{L. Wolfenstein, Phys. Rev. Lett. {\bf 13}
(1964) 562.} (1964), some small new interaction is postulated
to contribute at {\it lowest order} to the $\Delta S=2$
mass matrix, while the CKM phase $\delta=0$, yielding a scenario
where $\epsilon\neq 0$ but $\epsilon'=0$ to extremely good accuracy.

Within the standard model,
theoretical predictions for \epoe\ have varied a lot, and generally
shrunk, for two main reasons.  The first is that our expectation
for $m_t$, the top quark mass, has been steadily growing and
as $m_t$ grows, the amplitude of the gluonic penguin in Fig.~{\figf}
decreases.   In
the 1970's, 20 to 30 GeV was considered reasonable; towards the
end of the 80's estimates rose to around 100 GeV, with indirect
evidence from sources such as $B\bar B$ mixing measurements; now,
from LEP and Fermilab data we believe $m_t$ to be in the vicinity
of 170 GeV!  The second is that not only the contribution of the
gluonic penguin decreases as $m_t$ grows, but as $m_t$ enters
the 100 to 200 GeV range, the {\it electroweak} penguins
(like Fig.~{\figf}, only with a photon or a $Z$ replacing the gluon)
can no longer be neglected as they were originally.  In particular
the $Z$ penguin has significant and {\it cancelling} contributions.
A novel feature was thus that \epoe\ could actually pass through
zero, for large enough $m_t$ (around 200 GeV).
In Fig.~{\figg} I have shown some sample recent predictions
(from Buras and Harlander\Ref\Burasa{A. J. Buras and M. K. Harlander,
MPI-PAE/PTh 1/92, in the {\it Review Volume on Heavy Flavours},
eds. A. J. Buras and M. Lindner, Advanced Series on Directions
in High Energy Physics, World Scientific Publishing Co., Singapore.}
in 1992)
for \epoe\ versus $m_t$, in the two allowed quadrants for the
phase $\delta$,
for the parameter ranges
$0.09\le |V_{ub}|/|V_{cb}| \le 0.17$, $0.036\le |V_{cb}| \le 0.046$
(CKM matrix elements), $0.1$ GeV $\le \Lambda_{QCD} \le$ 0.3 GeV
(QCD scale), 0.5 $\le B_K\le$ 0.8 (bag factor), and
125 MeV $\le m_s \le$ 200 MeV (strange quark mass).

\epsffile[0 0 300 300]{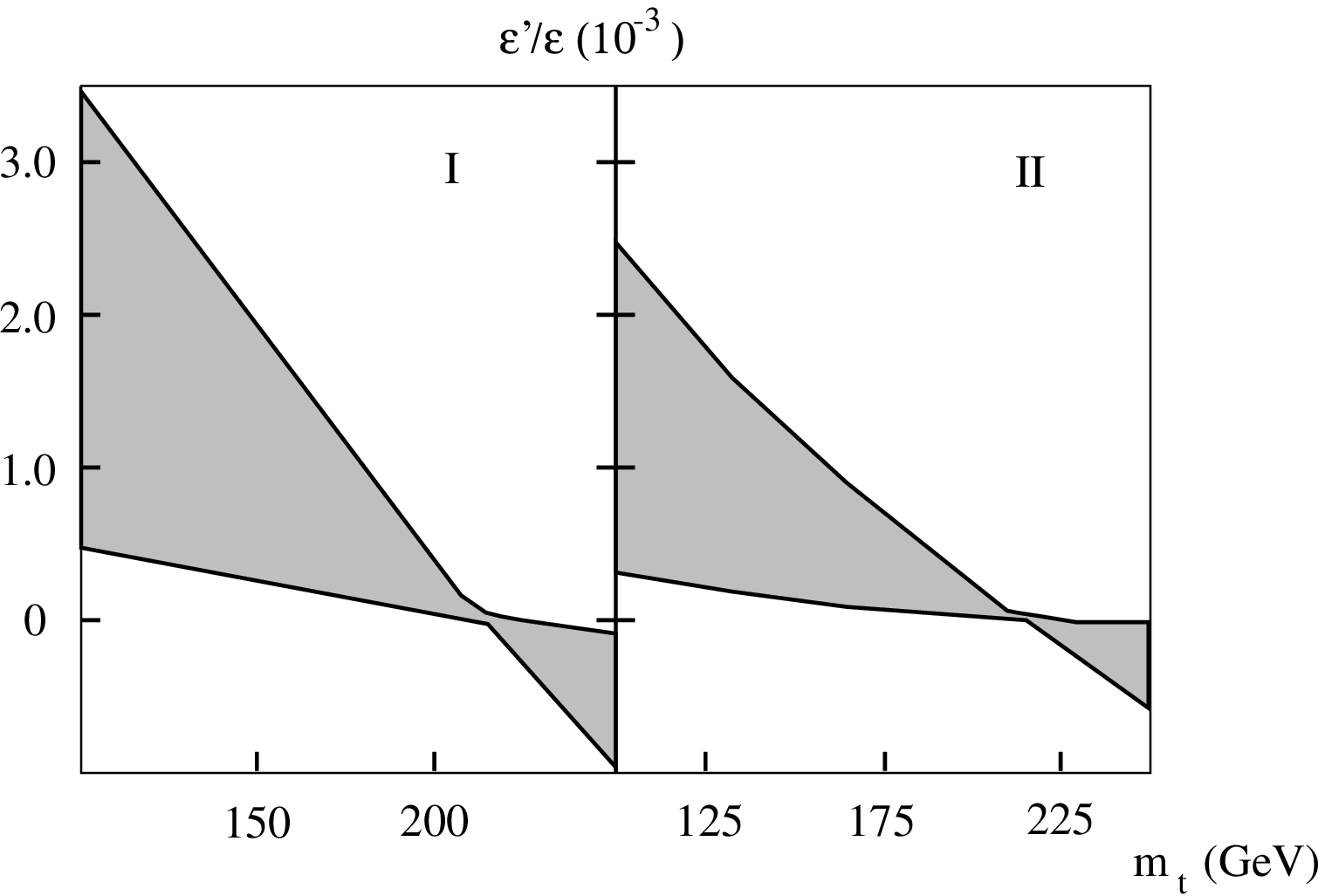}

{\bf Figure~{\figg}.} The upper and lower limits of \epoe\ in the
first (I) and second (II) quadrant of the CKM phase $\delta$
(from Ref. \Burasa).

Some of the historical progress of theoretical predictions,
which bears some resemblance to the evolution of the measurements,
is summarized in Fig.~{\fige}.
The first prediction, by Ellis, Gaillard and
Nanopoulos,\Ref\Egn{J. Ellis, M. K. Gaillard, and D. V. Nanopoulos,
Nucl. Phys. {\bf B109} (1976) 213.}
was quite small.  The second prediction
shown, of Gilman and Wise,\Ref\GWa{F. J. Gilman and M. B. Wise,
Phys. Lett. {\bf 83B} (1979) 83.} covered a range only starting
at $0.01$ and going all the way to $0.08$ for
$\epsilon'/\epsilon$.  The next prediction is a more
detailed calculation by
Gilman and Wise.\Ref\GWb{F. J. Gilman and M. B. Wise,
Phys. Rev. {\bf D20} (1979) 2392.}
The next two predictions are {\it lower limits} only, both
by Gilman and Hagelin.\Ref\GH{F. J. Gilman and J. S. Hagelin,
Phys. Lett. {\bf 126B} (1983) 111; {\bf 133B} (1983) 443.}
Neither bound is still valid, with current knowledge and a
large top mass.
The next rectangle shows the range of \epoe\ that was
considered reasonable\Ref\Wolf{L. Wolfenstein, in the {\it
Proceedings of the Second Lake Louise Winter Institute on New
Frontiers in Particle Physics} (Chateau Lake Louise, Canada,
15-21 February 1987).} in 1987, with a reasonably large $m_t$
(in the range of 100 GeV)
and $B\bar B$ mixing constraints taken into account.
Finally the last rectangle is representative of the predictions
in the last couple of years.  This is a collection: individual
preferred ranges may resemble the NA31 measurement or the
E731 measurement.
\REF\Burasb{A. J. Buras, M. Jamin and M. E. Lautenbacher,
Nucl. Phys. {\bf B408} (1993) 209.}
See for example Ref. \Burasb\ for a recent look at \epoe\ and
references to other work.
While \epoe\ at or below zero is in principle
allowed, it is not very favored, as it would require a very
heavy top quark (over 200 GeV).

\chapter{\epoe\ at a $\phi$-factory}
At a $\phi$-factory, $K$'s are produced {\it strongly} in pairs
in the reaction $e^+ e^-\rightarrow \gamma^* \rightarrow
\phi\rightarrow K^0 \bar K^0$ so that
$$ C(K^0 \bar K^0) = C(\phi) = C(\gamma) = -1.
\eqn\zxix$$
The initial state is thus
$${1\over \sqrt{2} } \left( \ket{K^0,\vec p}\ket{\bar K^0,-\vec p}
-\ket{\bar K^0,\vec p}\ket{K^0,-\vec p} \right)
\eqn\zxxa$$
which can be shown to be equal to
$${1\over \sqrt{2} }{1+|\epsilon|^2\over 1-\epsilon^2 }
\left( \ket{K_S,-\vec p} \ket{K_L,\vec p} -
 \ket{K_S,\vec p} \ket{K_L,-\vec p} \right).
\eqn\zxxb$$
Thus the $\phi$ decays to a pure
$K_S K_L$  state, with no admixture of $K_S K_S$ or $K_L K_L$.
Since $K_S$ and $K_L$ are weak eigenstates,
in vacuum the
pair of kaons remain in a state of pure $K_S$, $K_L$.
As a result \DAF\ benefits from `tagging':  by observing a clear
 signal for a $K_L$ ($K_S$) one can be sure that one has a
$K_S$ ($K_L$) as the other particle in the event, independent of
what this other particle decays to.   More specifically,
a `V' 20 to 180 cm from the interaction point signals a
$K_L\rightarrow \pi^+\pi^-\pi^0$, $\pi\mu\nu$, or $e\mu\nu$ with
no background ($<10^{-6}$) and is thus a perfect $K_S$ tag.
The efficiency for this tagging is $28\%$, since $K_L$'s
are very long lived (recall the mean path of 3.43 m)  and thus
do not all decay inside the detector).
  In other words, for
a well-defined sample of one-third of the all $K$'s, one can be
very sure one has a $K_S$.  Similarly, a $\pi^+\pi^-$ reconstructing
to within 2 cm of the interaction point tags the $K_L$.  The
background is two pion decays that come from $K_L$ instead
of $K_S$, and thus is of
the order of 8$\times 10^{-6}$ (the branching ratio for
$K_L$ going to two pions, times the probability that it does so,
so close to the interaction point).  Here the efficiency is
essentially
the percentage of $K_S$'s decaying to charged rather than neutral
pions, 2/3.  These efficiencies drop out {\it identically} from the
double ratio defined in Eq. \zxv, and thus produce no {\it systematic}
errors, only a loss in statistics.
 Another feature of $K$'s  at a $\phi$ factory is that they have a
very precise --- and small ($\beta =0.2$) --- momentum.

Systematic errors are beyond my scope,
but I can now at least demonstrate, on the statistical side,
the figures justifying the claimed accuracy for \epoe\ at \DAF.
In abbreviated form, the double ratio is
$${ N_L^\pm/N_S^\pm\over N_L^0/N_S^0}
\approx 1 + 6 {\rm Re} {\epsilon'\over \epsilon},
\eqn\zxxi$$
where each $N$ refers to the number of $K_{L,S}$ decaying to two
charged or neutral pions.  The $N_S$ will evidently be much
larger then the $N_L$; thus, the statistical errors coming from
them will be negligible compared to those coming from the $N_L$.
We thus have
$$ \delta\left( {\epsilon'\over \epsilon} \right)
= {1\over 6} \sqrt{(\Delta  N_L^\pm)^2 + (\Delta  N_L^0)^2}
=  {1\over 6} \sqrt{ {1\over N_L^\pm} + {1\over N_L^0} }
= {1\over 6} \sqrt{ {3\over 2 N^0_L}},
\eqn\zxxii$$
since by isospin symmetry there are twice as many charged two
pion decays as neutral two pion decays.
$N^0_L$ is given by the $\phi$ cross-section, times the integrated
luminosity, times the efficiency for $K_L$ tags, times the
$BR(\phi\rightarrow K_L K_S)$, times the $BR(K_L\rightarrow
\pi^0 \pi^0)$, times the number of $K_L$'s that are within the
fiducial volume (i.e., that are detectable):
$$N^0_L = 5 \mu{\rm b} \times 10^{10} \mu{\rm b}^{-1} \times
2/3 \times 0.34 \times 10^{-3} \times (1- e^{-150/350})
=4\times 10^6,
\eqn\zxxiii$$
which gives as claimed,
$$\delta\left( {\epsilon'\over \epsilon} \right)
= 1\times 10^{-4}.
\eqn\zxxiv$$

\chapter{Interferometry}
Defining $\eta_i=\langle\,f_i\sta{\kl}/\langle\,f_i\sta{\ks}$,
 $\Delta t = t_1 - t_2$, $ t = t_1 - t_2$,
$\Delta {\cal M} = {\cal M}_L - {\cal M}_S$,
${\cal M} = {\cal M}_L + {\cal M}_S$, and
${\cal M} = M_{L,S} - i \Gamma_{L,S}$,
the amplitude for decay to states $f_1$ at time $t_1$ and
$f_2$ at time $t_2$, without identification of $K_S$ or
$K_L$, is:
$$\eqalign{&
\langle\,f_1,\,t_1,\,{\bf p};\ f_2,\,t_2,\,{-\bf p}\sta{i}=
{1\over\sqrt2}{1+|\epsilon|^2 \over 1-\epsilon^2}\x\cr
&\kern-5mm\langle\,f_1\sta{\ks}\langle\,f_2\sta{\ks}e^{-
i{\cal M}t/2}\Big(\eta_1e^{i\Delta {\cal M}\Delta t/2}-
\eta_2e^{-i\Delta {\cal M}\Delta t/2}\Big).\cr
}\eqn\zxxv$$

The decay intensity $I(f_1,f_2,\Delta t=t_1-t_2)$ to final states $f_1$
and $f_2$ is obtained from eq. \zxxv\ above by integrating over all
$t_1,\ t_2$, with $\Delta t$ constant. For $\Delta t>0$:
$$\eqalign{&I(f_1,\ f_2;\ \Delta t)
={1\over2}\int_{\Delta t}^\infty|A(f_1,\ t_1;\ f_2,\ t_2)|^2\dt=\cr
&\qquad{1\over2\Gamma}|\langle f_1\sta{\ks}\langle f_2\sta{\ks}|^2
\Big(|\eta_1|^2e^{-\Gamma_L\Delta t}+|\eta_2|^2e^{-\Gamma_S\Delta t}\cr
&\qquad-2|\eta_1||\eta_2|e^{-\Gamma\Delta t/2}\cos(\Delta
m\Delta t+\f_1-\f_2)\Big),\cr  }\eqn\eqdec$$
with $\eta_i=A(\kl\to f_i)/A(\ks\to f_i)=|\eta_i|e^{i\f_i}$, exhibiting
interference terms sensitive to phase differences.

With the double ratio, \DAF\ complements future fixed target \epoe\
experiments, but for {\it interferometry}, it is unique.  It can
even test CPT conservation.  If one is completely general,
and does not assume CPT conservation, eqs. \ziv\ and \zv\
become
$$\sta{K_S} \propto \big((1+ \epsilon_\K + \delta_\K)\sta{\ko}+
(1-\epsilon_\K-\delta_\K)\sta{\kob}\big )/\sqrt2 $$
$$\sta{K_L}\propto \big ((1+\eps_{\K}-\delta_{\K})\sta{\ko}-
(1-\eps_{\K}+\delta_{\K})\sta{\kob}\big )/\sqrt2 .$$
In Fig.~{\figh} I show a sample interference pattern,
taken from Ref. \jlf, for the process shown in Fig.~{\figha}.
  The determination of \rep\ comes from the
difference in height of the two `shoulders.'  The large value
of \rep\ is only to make the difference visible in the graph;
KLOE will have a similar sensitivity to \rep\ from interference
as that discussed in the previous section.

\epsffile[0 0 365 365]{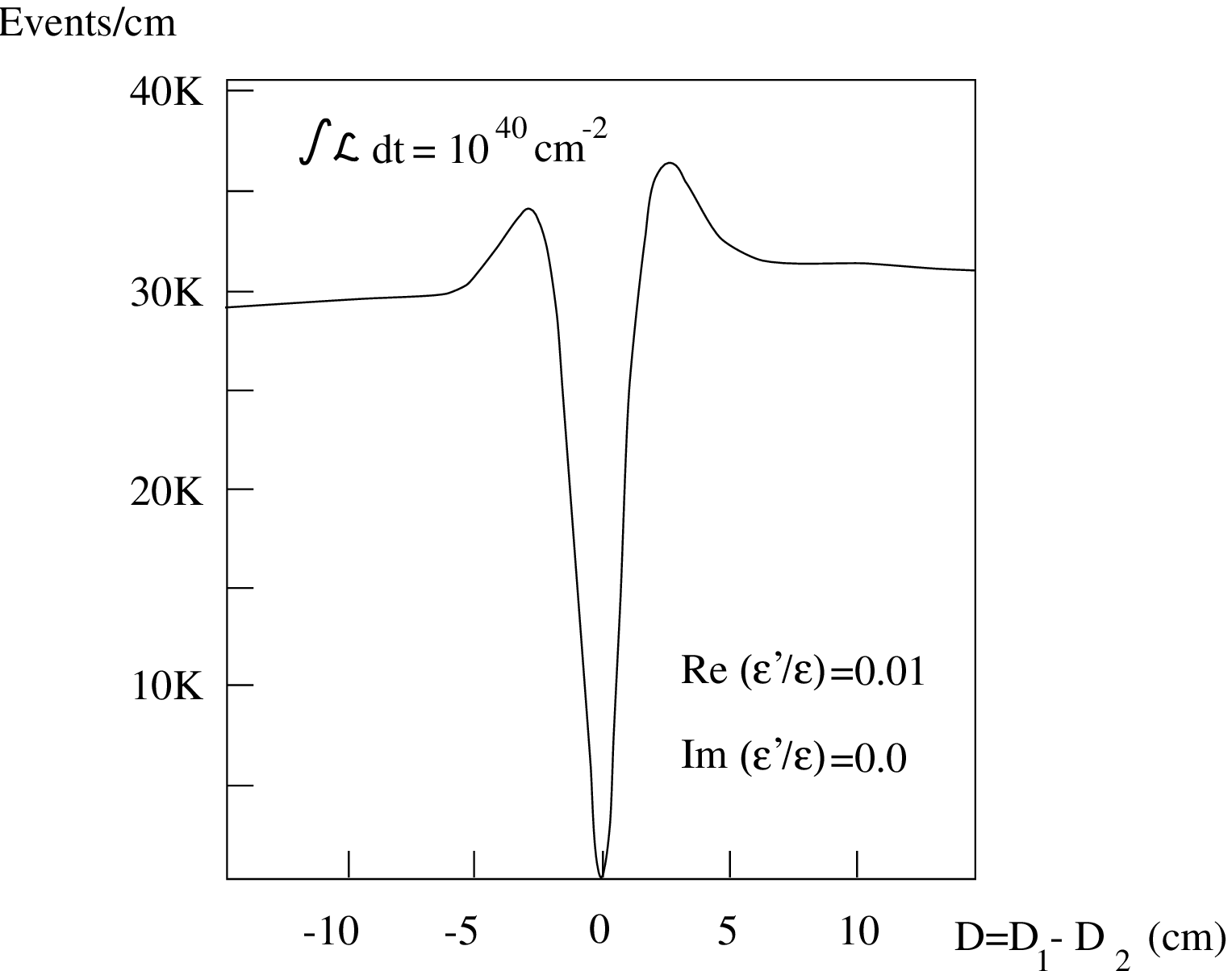}

\centerline{{\bf Figure~{\figh}.}  The interference pattern for
$f_1$=\pic,\  $f_2$=\pio\ \To\ \rep, \  \imp.  }

\epsffile[-50 0  165 165]{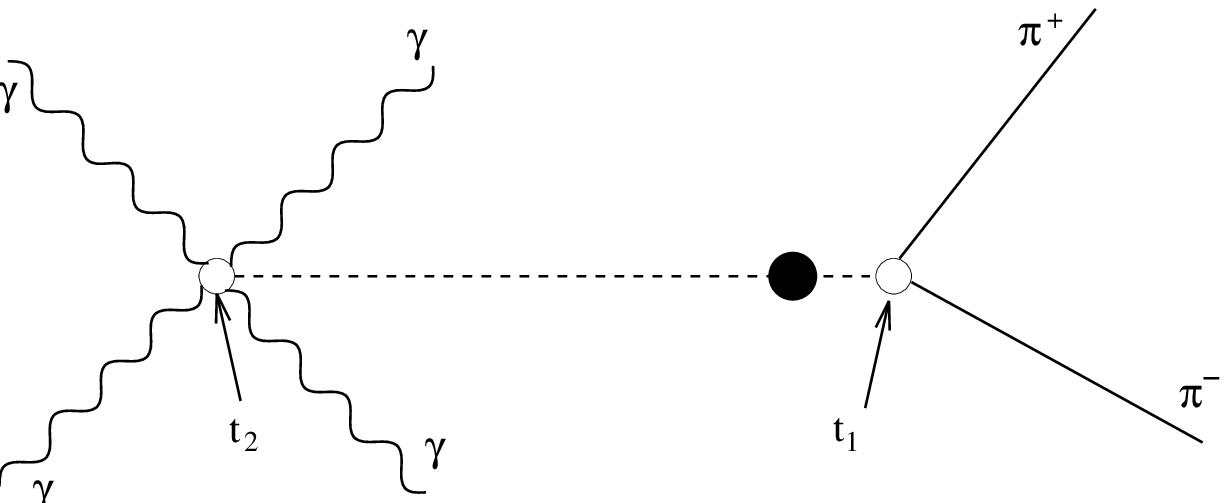}

\vskip .2cm
\centerline{{\bf Figure~{\figha}.} $\phi\rightarrow K_S K_L
\rightarrow f_1f_2$ with $f_1$=\pic,\  $f_2$=\pio.}

With various final states $f_1$ and $f_2$, all $K$ system parameters
can be measured independently.  Just a few  examples are:
\pointbegin $f_1= f_2$: $\Gamma_S$, $\Gamma_L$
and $\Delta m$ can be measured,
 since all the phases disappear. Rates can be measured
to \x10 improvement in accuracy and $\Delta m$ to \x2.
\point $f_1$=\pic,\ $f_2$=\pio:  \rep\ and \imp\ can be measured,
the former by concentrating on large time differences, the
latter for $|\Delta t|\le 5 \tau_s$.
\imp\ can be measured to accuracies of $10^{-3}$ (comparable
to other future experiments of the same epoch, but using a
completely different method).
\point $f_1=\pi^+\ell^-\nu$ and $f_2=\pi^- \ell^+\nu$:
the CPT--violation parameter $\delta_K$ can be measured,
the real part
by concentrating on large time difference regions; and the
imaginary part for $|\Delta t|\le 10 \tau_s$.
(See Ref. \jlf\ for a more complete list.)

\chapter{Other Searches for CP violation}
So far CP violation has only been seen in $K_L$ decays
($K_L \rightarrow \pi\pi$ and semileptonic decays).
\DAF\ can look for $K_S \rightarrow \pi^0\pi^0\pi^0$, the
counterpart to $K_L\rightarrow \pi\pi$.  The branching ratio (BR)
for this process is proportional to $\epsilon + \epsilon'_{000}$
where $\epsilon'_{000}$ is a quantity similar to $\epsilon'$,
signalling direct CP violation.  It is not as suppressed as the
normal $\epsilon'$, perhaps a factor of twenty less.
Nonetheless, as the expected BR is $2\times 10^{-9}$, the whole
signal will be at the 30 event level, and therefore there is here
only the chance to see CP violation in a new channel, not direct
CP violation.  The current limit on this BR is $3.7\times 10^{-5}$.
Another possibility is to look at the difference in rates between
$K_S\rightarrow \pi^+ l^- \nu$ and $K_S\rightarrow \pi^- l^+ \nu$,
which is expected to be $\sim 16\times 10^{-4}$ in one year's
running at \DAF, with an expected accuracy of $\sim 4\times 10^{-4}$.
Again this would be only a measurement of $\epsilon$, not $\epsilon'$,
but the observation for the first time of CP violation in two
new channels would be nonetheless of considerable interest.

CP violation can also be looked for in the decays of $K^{\pm}$.
Here there will be probably no signal, but limits will be greatly
improved.

\chapter{Conclusions}
Aside from being an excellent, dedicated environment for studying
CP violation, where \epoe\ will be determined to an accuracy
of $10^{-4}$ in one year's running, and {\it all} the $K$ system
parameters will be determined via interferometry, \DAF\ will also
be a
rich source of many other physics results.  \DAF\ will for example
be a unique source of pure $K_S$, thanks to tagging, providing
up to $10^{10}$ kaons per year, and measuring rare $K_S$ decay modes,
most of which have not been measured yet, down to the $10^{-8}$
or $10^{-9}$ level.  Rare charged $K$ decay modes will also be
studied.  \DAF\ will provide much input for CHPT, as Ulf Meissner
has discussed.
 It will help us to understand the enigmatic $f_0$;
it will be a place to study many rare decays, such as
$\phi\rightarrow\eta\gamma$, and $\eta$ decays;
to measure $\sigma(e^+e^-\rightarrow$hadrons) at energies
up to 1.5 GeV, which is necessary for the calculation of the
muon anomaly $a_\mu$; to study photon photon interactions;
and even to test the non-local character of quantum theory.
Finally, one will be able to understand better the strange sea
quark content of the nucleon and study hypernuclei by looking at
 kaon-nucleon scattering.  I conclude by listing in Table 2 some
of the improvements that may be made in various rare decay
BR limits.
\vskip .2in
\begintable
Decay mode     | To date | Limits that can be achieved
at \DAF\ \cr
$\eta\rightarrow 3 \gamma$ | BR $< 5\times 10^{-4}$ |
$ 1.4 \times 10 ^{-8} $ \cr
$\eta\rightarrow \omega \gamma$ | BR $< 5\times 10^{-2}$  |
  $ 10 ^{-9}$  \cr
$\phi\rightarrow \rho \gamma$ | BR $< 2\times 10^{-2}$  |
  $ 10 ^{-9}$  \cr
$\phi\rightarrow \pi^+ \pi^- \gamma$ | BR $< 7\times 10^{-3}$  |
  $ 10 ^{-9}$  \cr
$\eta\rightarrow \pi^0 e^+ e^-$ | BR $< 4\times 10^{-5}$  |
$ 1.4 \times 10 ^{-8}$  \cr
$\eta\rightarrow \pi^0 \mu^+ \mu^-$ | BR $< 5\times 10^{-6}$  |
$ 1.4 \times 10 ^{-8}$   \endtable
\centerline{{\bf Table 2.} Rare decays at \DAF.}
\vskip .2in

\chapter{Acknowledgements}
I would like to thank my colleagues at PSI for organizing such a enjoyable
summer school.  It is also a pleasure to thank Juliet Lee-Franzini and
Paolo Franzini for many enlightening, indispensable discussions and
a careful reading of this manuscript.

\par \penalty-400 \vskip\chapterskip
   \spacecheck\referenceminspace \immediate\closeout\referencewrite
   \referenceopenfalse
   \line{\fourteenrm\hfil REFERENCES\hfil}\vskip\headskip
   \input referenc.texauxil
   
\bye